\documentclass[12pt, preprint]{aastex}
\usepackage{graphicx}
\usepackage{lscape}
\newcommand{\kms}{km\,s$^{\rm -1}$}

\newcommand{\co}{$^{12}$CO}                             
\newcommand{\xco}{$^{13}$CO}                            
\newcommand{\xxco}{C$^{18}$O}                           
\begin{document}
\title{A Possible Extension of the Scutum-Centaurus Arm into the Outer Second Quadrant}
\author{Yan ~Sun$^{1, 2, 3}$, Ye Xu$^{1}$, Ji Yang$^{1}$, Fa Cheng
Li$^{1}$, Xin Yu Du$^{1}$, Shao Bo Zhang$^{1}$, Xin Zhou$^{1}$}
\altaffiltext{1} {Purple Mountain Observatory, Chinese Academy of
Sciences, Nanjing 210008, China; yansun@pmo.ac.cn}
\altaffiltext{2} {Graduate University of the Chinese Academy of Sciences,
19A Yuquan Road, Shijingshan District, Beijing 100049, China}
\altaffiltext{3} {Key Laboratory of Radio Astronomy, Chinese Academy of
Sciences, China}
\begin{abstract}
Combining HI data from the Canadian Galactic Plane Survey~(CGPS) and CO data
from the Milky Way Imaging Scroll Painting (MWISP) project, we have identified a
new segment of a spiral arm between Galactocentric radii of 15 and 19~kpc
that apparently lies beyond the Outer Arm in the second Galactic quadrant. Over
most of its length, the arm is 400-600 pc thick in $z$. The new arm appears
to be the extension of the distant arm recently discovered by \citet{dame2011}
as well as the Scutum$-$Centaurus Arm into the outer second quadrant.
Our current survey identified a total of 72 molecular clouds with masses on the order of
10$^2$-10$^4$M$_{\sun}$ that probably lie in the new arm. When all of the
available data from the CO molecular clouds are fit, the best$-$fitting spiral
model gives a pitch angle of 9.3$^{\circ}$ $\pm$0.7$^{\circ}$.
\end{abstract}


\section{Introduction}
Our knowledge of the outer Galaxy to date is largely
dependent on HI emission surveys. The 21$-$cm line observations outline a
spiral structure out to at least 25~kpc, implying a minimum radius for the
gas disk~\citep{levine2006}. Through the International Galactic Plane
Survey (IGPS), a widespread presence of cool HI~(traced by HI absorption) was
found between Galactocentric radii of 12 and 25~kpc~\citep{strasser2007}.
It was not until 2011, however, that a segment of a spiral arm at radii of $\sim$15~kpc
was identified by \citet{dame2011} beyond the Outer Arm in the first Galactic quadrant.
This was accomplished by using the Leiden/Argentine/Bonn~(LAB) 21$-$cm survey
~\citep{kalberla2005}, and new CO observations with the 1.2~m telescope at
the Harvard-Smithsonian Center for Astrophysics~(CfA).

Because of the far distances involved, high-sensitivity CO observations are
essential for delineating spiral structures in the extreme outer
Galaxy. \citet{digel94} were the first to detect molecular clouds in this region. 
Due to the relatively low sensitivity of the Five College Radio Astronomy Observatory~(FCRAO)
CO survey~\citep{heyer98}, only a few bright molecular clouds were identified there~\citep{kerton2003}.
Until now, no coherent, large-scale structures beyond the Outer Arm
have been identified in the second quadrant.

We have carried out large-scale CO mapping in the second quadrant of the Galactic plane,
which is part of the Milky Way Imaging Scroll Painting (MWISP) Project\footnote{\url{
http://www.radioast.nsdc.cn/yhhjindex.php}}. One of the goals of this project
is to probe the spiral structures further than the Outer Arm.
This mapping has completely covered the regions between 120$^{\circ} \leq l \leq
144^{\circ}$ and -2.25$^{\circ} \leq b \leq 3.75^{\circ}$,
and between 144$^{\circ} \leq l \leq 150^{\circ}$ and -3.75$^{\circ} \leq b \leq 3.75^{\circ}$.
In the longitude range of 100$^{\circ}$ to 120$^{\circ}$, the observations so far
were guided by the latitude distribution of extreme outer molecular clouds
previously reported by \citet{brunt2003} and \citet{kerton2003}, rather
than an unbiased survey.

\section{CO observations and Archival Data of atomic hydrogen}
\subsection{CO observations}
The observations were conducted during 2011 and 2014 using the 13.7-m telescope
of the Purple Mountain Observatory~(PMO) in Delingha, China.
The receiver was the newly installed Superconducting Spectroscopic Array
Receiver~\citep[SSAR;][]{shan2012}. All observations were carried out in ``on-the-fly''
(OTF) mode. The \co,~ \xco~ and \xxco~ lines were observed simultaneously.

At 115~GHz, the main beam width was about 52$\arcsec$, and the main beam
efficiencies~($\eta_{mb}$) were 0.46 for \co~ and 0.43 for \xco~ and \xxco.
The typical rms noise level was $\sim$ 0.5~K for \co\ and
0.3~K for \xco~ and \xxco, corresponding to a channel width of 0.16~\kms.
All of the spectral data were processed using the GILDAS package.
It should be noted that any results presented in the figures and
tables are on the brightness temperature scale~(T$_R^*$), corrected
for beam efficiencies using T$_{mb}$=T$_A^*$/$\eta_{mb}$.
Details on the observations, data collection, and data processing
steps will be presented in Sun et al. (2014, in prep.).

\subsection{Archival Data of atomic hydrogen}
The 21-cm line data were retrieved from the Canadian Galactic Plane
Survey~\citep[CGPS;][]{taylor2003}, covering Galactic longitudes from
$l$ = 63$^{\circ}$ to 175$^{\circ}$ and latitudes from $b$ = -3.5$^{\circ}$
to +5.5$^{\circ}$, and a latitude extension to $b$ = +17.5$^{\circ}$
between $l$ = 100$^{\circ}$ to 116.5$^{\circ}$. The velocity coverage
of the data is in the range of -153 to 40~\kms~ with a channel separation
of 0.82~\kms. The survey has a spatial resolution of 58$\arcsec$ which
is comparable to our CO observations.

\section{Results and Discussion}
We created $b-v$ maps of the HI emission in the CGPS~(66$^{\circ}$$<l<$150$^{\circ}$)
and the CO emission in the MWISP~(100$^{\circ}$$<l<$150$^{\circ}$), by obtaining
average emissions over a 5$^{\circ}$ wide window in longitude.
The data of the CGPS in the range of 150$^{\circ}<l<$175$^{\circ}$ were excluded
from this analysis because of the degeneracy of the radial velocity.
The Outer Arm has been widely accepted as the outermost arm in the second quadrant
~\citep{russeil2003,russeil2007}. However, all $b-v$ maps reveal a clear and distinct feature
distributed beyond the Outer Arm at a maximum negative velocity.
This outermost feature varies in latitude and velocity.
In Figure 1, we show one of these $b-v$ distributions from both the CO~(image) and the HI~(contours)
emissions in longitude from 127$^{\circ}$ to 132$^{\circ}$.
The Outer Arm and the outermost feature are clearly seen as strong, separate features.

To highlight the outermost feature, a $l-v$ map of integrated HI emission over
a window that follows the feature in latitude is shown in Figure~2a.
The high-mass star forming regions~(HMSFRs) assigned to the Outer
and Perseus Arms are marked with squares and diamonds, respectively~\citep{reid2014}.
Similarly, Figure~2b focuses on the outermost feature, obtained by
integrating the HI emission over a window that follows the feature in velocity.
It is worth noting that at large negative velocities, the HI emission
forms an arc from $l\sim$75$^{\circ}$, $b\sim$3$^{\circ}$,
$v\sim$-110~\kms, through $l\sim$110$^{\circ}$, $b\sim$3.5$^{\circ}$, $v\sim$-105~\kms,
to $l\sim$150$^{\circ}$, $b\sim$-2$^{\circ}$, $v\sim$-70~\kms,
running roughly parallel to the locus of the Outer Arm but shifted by 20-30~\kms~
to more negative velocities. This suggests that it may be a segment of a major
spiral arm since spiral arms have long been recognized
as presenting coherent arcs and loops in Galactic $l-v$ plots
of atomic and molecular emissions~\citep{weaver74,dame86,dame2008,dame2011}.
The present feature was largely overlooked in the past since only short segments
of it appear in l-v diagrams at any specific latitude or even in an l-v diagram
integrated over all latitudes covered by the CGPS~\citep[e.g., in Figure 3 of][]{strasser2007}.

Before giving an explicit claim of a new spiral arm, we investigated the
characteristics of the outermost feature in more detail.
So far a total of 72 molecular clouds that are probably attributed to
the outermost feature were detected by MWISP. These 72 clouds are
marked with circles in Figures 2 and 3 and also
summarized in Table~1. More than 10 clouds show clear
\xco~ detections while many of other clouds were only marginally detected.
However, none of them show \xxco~ detections. The isotope abundance ratios 
in the extreme outer Galaxy will be studied in the forthcoming deep observations.
Actually, all known clouds in the extreme outer Galaxy reported by \citet{digel94}
and \citet{kerton2003} were detected by the present survey.
We labelled all these known clouds in column (1) of Table 1.
Note that the clouds reported by \citet{digel94} in the higher
longitude range of 145$^{\circ}<l<$151$^{\circ}$ are not included in
Table 1, since these clouds are partially blended in velocity with
emission at less negative velocities. Besides, they might lie in the
Outer arm or in the inter-arm on the assumption of the rotation curve of
\citet{reid2014}.

Most of the clouds are newly discovered, which suggests
that we may still miss some molecular clouds in the interval
100$^{\circ}<l<$120$^{\circ}$ due to the narrow latitude coverage.
The derived masses, mainly on the order of 10$^{\rm 3}$M$_{\sun}$ to 10$^{\rm 4}$M$_{\sun}$,
fall in the typical range for the masses of molecular clouds
that are usually confined to spiral arms. Note that the mass is
a lower limit, since the adopted CO-to-H$_2$ X factor
1.8$\times$10$^{20}$ cm$^{-2}$ is measured in the solar neighborhood.
However, a recent study suggested that the X factor rises by a
factor of 2$-$3 between the solar circle and the circle at
$R_{\rm Gal}$=14~kpc~\citep{abdo2010}. In addition,
the beam dilution effects lower the signal-to-noise ratio
and hence also reduce the derived masses.
We have selected one example with the largest negative velocities we
recently detected~(molecular cloud 24 marked with cross)
and presented the velocity-integrated
intensity and spectrum of the peak emission in Figure 2b.
Although these molecular clouds are more sparsely distributed than
the atomic gas, both explicitly trace the same coherent, large-scale
structure, i.e., the new arm well.

The warp of the Galactic disk is obvious~(Figure 2b), which exhibits a
clear positive warp between $l$ = 66$^{\circ}$ and 135$^{\circ}$,
and a negative warp between $l$ = 135$^{\circ}$ to 150$^{\circ}$.
Assuming a mean distance of 12~Kpc, the new arm with
thickness in $z$ of about 2$^{\circ}$ to 3$^{\circ}$
correspond to about 400 to 600~pc. Generally, the new
arm is more severely warped and thicker than the Outer
Arm. This is in good agreement with the theory for both
external spiral galaxies and our Galaxy, which suggests that the 
warping and thicknesses of the disks increase with increasing distance 
from the galactic center~\citep{wouterloot90,russeil2003,bottema96}.
Moreover, the new arm is distinct from the Outer Arm with a spatial
separation of about 3$-$4~kpc, which is consistent with the
typical span of the spiral arms outside the Perseus Arm~\citep{reid2014,vallee2014}.

Locations of all available molecular clouds in the far outer Galaxy are compared
with the spiral arm model in Figure~3, superposed on: (a) CO $l-v$ diagram from
\citet{dame2001}, and (b) an artist's conception of the Milky Way~(R. Hurt:
NASA/JPL-Caltech/SSC). The artist's image has been scaled to place the HMSFRs
in the spiral arms. These clouds are indicated by filled stars, triangles and
circles, which cover Galactic longitudes from $l$=13$^{\circ}$ to 55$^{\circ}$
detected by \citet{dame2011}, and from $l$=100$^{\circ}$ to 150$^{\circ}$ detected by us.
The squares indicate the locations of the HMSFRs associated with the Outer Arm~\citep{reid2014}.
The white dashed line in Figure 3 is a log spiral that was fit to the 
Scutum-Centaurus arm in the inner Galaxy~\citep{vallee2008,vallee2014}.
Generally, our data are in agreement with the far-extension of the Scutum-Centaurus Arm
in the second quadrant. However, confirmation of this hypothesis would be a challenge as
suggested by \citet{dame2011}.

We attempted to fit the distribution of these outermost clouds, adopting a
log$-$periodic spiral model defined by ${\rm ln} (R/R_{\rm ref}) = -(\beta - \beta_{\rm ref}) tan\psi$,
where $R$ is the Galactocentric radius at a Galactocentric azimuth $\beta$ for
an arm with a radius $R_{\rm ref}$ at reference azimuth $\beta_{\rm ref}$ and pitch angle $\psi$.
The Bayesian Markov chain Monte Carlo (McMC) procedure was adopted  to estimate
the parameters $R_{\rm ref}$ and $\psi$ (see paper of \citet{reid2014} for more details).
The best$-$fitting spiral model for the CO data on the interval
120$^{\circ}\leq l\leq$150$^{\circ}$ is $R_{\rm ref}$ = 17.8$\pm$1.0~kpc at
$\beta_{\rm ref}$ = 27.0$^{\circ}$ and $\psi$= 10.8$^{\circ}$ $\pm$
7.6$^{\circ}$~(indicated by the red dashed line in Figure 3). The best$-$fitting
spiral model for all of the available CO data including clouds detected by 
\citet{dame2011} is $R_{\rm ref}$ = 16.1$\pm$1.0~kpc
at $\beta_{\rm ref}$ = 43.9$^{\circ}$ and $\psi$= 9.3$^{\circ}$ $\pm$
0.7$^{\circ}$~(indicated by the cyan solid line in Figure 3).
The uncertainties give a 68\% confidence range.

The consistent fitting results confirm that the new arm can be interpreted
as the far-extension of the distant arm recently discovered by \citet{dame2011}.
Apparently, both models fit the data well over a wide longitude range, except in
$l$=100$^{\circ}$ to 116$^{\circ}$, where the model deviates from the data.
Such departures might be caused by the kinematic distance uncertainty.
Indeed, any systematic velocity departures can lead to distance uncertainty.
The velocity departures, including both positive and negative values have
already been observed, e.g., in the Perseus Arm between 90$^{\circ}$ and
150$^{\circ}$ and the Sagittarius-Carina Arm~\citep{russeil2003}.
Another possibility is the bias introduced by the incompleteness of
the sample due to the narrow latitude coverage in this longitude interval,
as mentioned above. The parallax measurements and MWISP projects in
progress will be required to solve the problem and determine the new
arm more accurately.

We are grateful to all of the staff of the 14-m telescope of the PMO for their
dedicated assistance. We would like to thank Dr. Mark Reid for using of his
fitting procedure. We thank Dr. Yuan-Wei Wu for his useful discussions.
We also thank the anonymous referee for very helpful suggestions
and comments that help improved the paper.
Research for this project is supported by the National Natural Science
Foundation of China (grant Nos. 11003046, 11233007, 11133008, 11403104),
the Strategic Priority Research Program of the Chinese Academy
of Sciences (grant No. XDB09000000), and the Key
Laboratory for Radio Astronomy, CAS.


\setlength{\tabcolsep}{0.04in}
\begin{deluxetable}{lcccccccccccc}
\tabletypesize{\tiny}
\tablewidth{0pt} \tablecaption{Parameters of molecular clouds derived by CO(1-0).\label{tab:detect}}
\tablehead{Number& $l$ & $b$ & $V_{\rm lsr}$ & $T_{\rm peak}$ & $\Delta$$V$ & $W_{\rm CO}$ & area & $d$ & $R$ & $Z$ scale& radius& Mass \\
    & ($\arcdeg$) &($\arcdeg$)&(\kms) & (K) & (\kms) & (K.km$\,$s$^{-1}$) &(arcmin$^2$) &(kpc)& (kpc) & (kpc) & (pc) & (10$^3$M$_\sun$)  \\
      (1)&(2)&(3)&(4)&(5)&(6)&(7)&(8)&(9)&(10)&(11)&(12)&(13)}
\startdata
  1&101.992 &   3.016&  -101.0 &  3.0&   3.3&  10.4&   39.4&    9.8 &   14.2 &    0.5 &  10.1& 13.0 \\
  2&102.092 &   2.776&  -102.9 &  3.5&   2.4&   8.9&   13.4&   10.1 &   14.4 &    0.5 &   5.9&  3.9 \\
  3&102.375 &   2.733&  -102.6 &  4.6&   1.3&   6.2&   29.0&   10.0 &   14.4 &    0.5 &   8.8&  5.9 \\
  4&103.042 &   2.475&  -106.4 &  2.5&   1.2&   3.3&    3.2&   10.5 &   14.8 &    0.5 &   2.8&  0.3 \\
  5&103.458 &   3.300&  -108.1 &  3.0&   0.9&   2.8&    3.7&   10.7 &   15.0 &    0.6 &   3.1&  0.3 \\
  6&103.729 &   2.867&  -100.5 &  3.2&   1.8&   6.1&    6.5&    9.7 &   14.2 &    0.5 &   3.9&  1.1 \\
  7*&104.983 &   3.317&  -102.7 &  5.8&   1.9&  11.4&   36.2&    9.9 &   14.5 &    0.6 &   9.7& 13.1 \\
  8*&105.242 &   3.025&  -100.9 &  4.0&   1.7&   7.5&   37.5&    9.6 &   14.3 &    0.5 &   9.6&  8.5 \\
  9&105.283 &   3.175&  -106.5 &  3.9&   2.0&   8.1&    5.6&   10.4 &   15.0 &    0.6 &   3.8&  1.5 \\
 10&106.417 &   3.925&  -112.0 &  1.5&   1.1&   3.2&   18.3&   11.1 &   15.7 &    0.8 &   7.6&  2.3 \\
 11*&107.725 &   2.933&  -102.1 &  2.5&   1.8&   4.8&   27.5&    9.7 &   14.6 &    0.5 &   8.2&  4.0 \\
 12 &107.900 &   1.908&  -102.5 &  1.8&   2.4&   4.5&   10.9&    9.7 &   14.7 &    0.3 &   5.1&  1.5 \\
 13 &109.200 &   2.283&  -103.0 &  4.3&   0.7&   3.4&    5.9&    9.8 &   14.8 &    0.4 &   3.7&  0.6 \\
 14* &109.292 &   2.083&  -101.2 &  5.2&   2.1&  11.5&   22.1&    9.5 &   14.6 &    0.3 &   7.3&  7.5 \\
 15* &109.375 &   2.642&   -98.6 &  3.6&   1.5&   5.9&   11.9&    9.2 &   14.3 &    0.4 &   5.1&  1.9 \\
 16* &109.500 &   2.608&   -99.6 &  4.5&   1.5&   7.4&   55.8&    9.3 &   14.5 &    0.4 &  11.3& 11.7 \\
 17* &109.642 &   2.700&   -98.8 &  4.8&   1.8&   9.1&   40.8&    9.2 &   14.4 &    0.4 &   9.6& 10.3 \\
 18* &109.790 &   2.717&   -99.2 &  5.8&   1.7&  10.3&   21.9&    9.2 &   14.4 &    0.4 &   7.0&  6.2 \\
 19 &110.025 &   3.083&   -98.1 &  3.3&   0.8&   2.9&   22.5&    9.1 &   14.3 &    0.5 &   7.0&  1.7 \\
 20 &110.167 &   2.783&   -98.3 &  3.2&   1.0&   3.3&    8.1&    9.1 &   14.4 &    0.4 &   4.1&  0.7 \\
 21* &114.342 &   0.781&  -100.9 &  7.9&   2.1&  18.0&  113.7&    9.5 &   15.0 &    0.1 &  16.5& 60.5 \\
 22 &115.992 &   1.250&  -115.8 &  2.4&   1.1&   2.6&    6.2&   11.9 &   17.3 &    0.3 &   4.7&  0.7 \\
 23* &116.725 &   3.542&  -107.6 &  5.5&   1.9&  11.3&   22.8&   10.5 &   16.1 &    0.6 &   8.1&  9.2 \\
 24 &117.367 &   1.700&  -120.8 &  2.1&   2.3&   5.1&   83.7&   13.0 &   18.4 &    0.4 &  19.5& 23.8 \\
 25* &117.576 &   3.950&  -106.0 &  5.9&   1.2&   7.7&   18.7&   10.3 &   16.0 &    0.7 &   7.2&  4.9 \\
 26* &118.143 &   3.417&  -106.7 &  2.8&   1.6&   4.9&   18.1&   10.4 &   16.2 &    0.6 &   7.2&  3.1 \\
 27 &120.925 &   2.758&  -103.3 &  2.8&   1.6&   4.8&    6.3&   10.1 &   16.1 &    0.5 &   4.0&  0.9 \\
 28 &121.375 &   2.708&  -103.6 &  2.3&   0.9&   2.4&   22.2&   10.2 &   16.2 &    0.5 &   7.8&  1.8 \\
 29* &121.675 &   2.042&  -101.5 &  2.6&   1.4&   3.9&   32.7&    9.9 &   16.0 &    0.4 &   9.2&  4.1 \\
 30* &121.817 &   3.052&  -103.9 &  5.6&   2.3&  13.7&   78.7&   10.3 &   16.4 &    0.5 &  14.9& 37.6 \\
 31 &121.967 &   1.842&  -102.8 &  2.5&   1.1&   3.0&   25.9&   10.1 &   16.2 &    0.3 &   8.4&  2.6 \\
 32* &122.375 &   1.775&  -102.5 &  3.0&   1.2&   3.9&   70.0&   10.1 &   16.3 &    0.3 &  13.9&  9.2 \\
 33 &122.492 &   2.592&  -105.1 &  2.4&   1.4&   3.5&   31.5&   10.6 &   16.7 &    0.5 &   9.7&  4.0 \\
 34* &122.775 &   2.522&  -107.2 &  5.5&   2.2&  12.8&   91.8&   11.0 &   17.1 &    0.5 &  17.3& 47.0 \\
 35* &123.367 &   1.659&  -103.2 &  2.6&   2.3&   6.5&   66.0&   10.4 &   16.5 &    0.3 &  13.8& 15.2 \\
 36 &123.650 &   2.833&  -104.9 &  3.1&   1.1&   3.6&   35.5&   10.7 &   16.9 &    0.5 &  10.4&  4.8 \\
 37 &123.925 &   3.158&  -105.4 &  2.9&   1.1&   3.2&   13.6&   10.8 &   17.0 &    0.6 &   6.4&  1.6 \\
 38 &124.292 &   3.000&  -109.8 &  1.9&   1.8&   3.6&   15.1&   11.7 &   17.9 &    0.6 &   7.4&  2.4 \\
 39 &124.342 &   3.367&  -104.9 &  1.8&   1.4&   2.8&    8.7&   10.8 &   17.0 &    0.6 &   5.1&  0.9 \\
 40 &124.525 &   3.400&  -103.9 &  2.7&   0.8&   2.3&   15.3&   10.6 &   16.8 &    0.6 &   6.7&  1.3 \\
 41 &125.075 &   2.375&  -106.9 &  1.7&   2.1&   3.9&   41.4&   11.3 &   17.5 &    0.5 &  11.9&  6.8 \\
 42 &127.150 &   2.583&  -101.5 &  2.1&   1.4&   3.1&   35.0&   10.6 &   17.0 &    0.5 &  10.2&  4.0 \\
 43* &127.875 &   2.167&   -99.6 &  2.4&   2.0&   4.9&   81.0&   10.4 &   16.9 &    0.4 &  15.3& 14.1 \\
 44 &128.633 &   2.833&  -102.4 &  2.2&   0.9&   2.2&    8.0&   11.1 &   17.6 &    0.5 &   5.0&  0.7 \\
 45 &130.050 &   2.108&  -102.5 &  3.6&   0.9&   3.4&   18.6&   11.4 &   18.0 &    0.4 &   8.0&  2.7 \\
 46$\dagger$ &131.016 &   1.524&  -102.2 &  2.5&   2.7&   7.2&   33.8&   11.6 &   18.3 &    0.3 &  11.0& 10.8 \\
 47$\dagger$* &131.157 &   1.390&  -100.6 &  3.8&   2.2&   8.6&   15.9&   11.3 &   18.0 &    0.3 &   7.3&  5.6 \\
 48 &131.575 &   1.300&  -104.5 &  1.9&   0.8&   1.6&    6.1&   12.4 &   19.0 &    0.3 &   4.8&  0.5 \\
 49 &132.792 &  -0.425&   -97.1 &  1.8&   2.1&   3.9&   10.2&   11.0 &   17.8 &   -0.1 &   5.6&  1.5 \\
 50 &133.817 &  -0.758&   -94.7 &  4.6&   1.1&   5.4&   31.3&   10.7 &   17.6 &   -0.1 &   9.8&  6.3 \\
 51 &133.825 &  -0.017&   -91.2 &  1.7&   1.0&   1.8&    3.0&   10.0 &   16.9 &   -0.0 &   2.6&  0.2 \\
 52 &136.859 &   1.557&   -90.6 &  2.3&   1.5&   3.5&   16.3&   10.7 &   17.8 &    0.3 &   7.0&  2.1 \\
 53* &137.283 &  -1.150&  -101.6 &  2.7&   2.4&   6.9&   13.2&   14.0 &   20.9 &   -0.3 &   8.2&  5.7 \\
 54 &137.517 &  -1.267&   -98.4 &  1.7&   1.8&   3.2&   10.5&   13.1 &   20.1 &   -0.3 &   6.8&  1.8 \\
 55 &137.617 &  -1.225&   -98.7 &  3.4&   1.5&   5.6&   10.6&   13.2 &   20.2 &   -0.3 &   6.9&  3.3 \\
 56$\dagger$* &137.759 &  -0.983&  -103.0 &  5.9&   2.8&  17.6&   41.0&   14.7 &   21.7 &   -0.3 &  15.4& 51.4 \\
 57$\dagger$* &137.775 &  -1.067&  -102.1 &  5.0&   2.3&  12.2&   43.0&   14.4 &   21.4 &   -0.3 &  15.4& 35.9 \\
 58 &138.373 &  -0.850&   -94.4 &  1.9&   1.6&   3.2&    8.7&   12.3 &   19.4 &   -0.2 &   5.8&  1.3 \\
 59 &139.116 &  -1.475&   -96.8 &  4.9&   1.3&   6.7&    4.6&   13.4 &   20.5 &   -0.3 &   4.4&  1.6 \\
 60 &139.850 &   0.368&   -88.9 &  1.9&   1.4&   2.9&    5.2&   11.4 &   18.6 &    0.1 &   4.0&  0.6 \\
 61 &140.183 &   0.258&   -88.3 &  2.0&   0.8&   1.8&    8.1&   11.4 &   18.6 &    0.1 &   5.1&  0.6 \\
 62 &140.700 &   0.150&   -86.9 &  1.9&   1.2&   2.5&    8.0&   11.2 &   18.5 &    0.0 &   5.0&  0.8 \\
 63 &141.083 &   0.425&   -85.0 &  1.3&   1.9&   2.6&    8.3&   10.8 &   18.2 &    0.1 &   5.0&  0.8 \\
 64 &142.167 &   0.267&   -85.8 &  1.6&   1.3&   2.3&    9.5&   11.6 &   18.9 &    0.1 &   5.7&  0.9 \\
 65$\dagger$ &142.642 &   0.317&   -81.1 &  2.0&   1.0&   2.0&   49.0&   10.5 &   17.9 &    0.1 &  12.0&  3.5 \\
 66 &143.325 &   1.775&   -79.9 &  2.7&   1.1&   3.1&   17.9&   10.4 &   17.9 &    0.3 &   7.1&  1.9 \\
 67 &143.533 &   0.492&   -76.7 &  4.0&   1.6&   6.5&   12.8&    9.7 &   17.2 &    0.1 &   5.6&  2.5 \\
 68 &144.167 &   0.817&   -74.6 &  3.0&   1.4&   4.6&    6.3&    9.4 &   16.9 &    0.1 &   3.7&  0.8 \\
 69 &145.208 &  -0.392&   -74.0 &  4.8&   1.7&   8.8&    7.5&    9.7 &   17.2 &   -0.1 &   4.2&  1.9 \\
 70 &145.808 &  -1.817&   -79.6 &  1.8&   1.8&   3.6&   17.1&   11.6 &   19.2 &   -0.4 &   7.8&  2.7 \\
 71 &145.850 &  -1.717&   -78.6 &  2.5&   1.7&   4.4&   60.0&   11.3 &   18.9 &   -0.3 &  14.3& 11.2 \\
 72 &146.059 &  -1.650&   -77.2 &  4.9&   1.7&   8.7&   17.6&   11.0 &   18.6 &   -0.3 &   7.5&  6.0 \\
\enddata
\vspace{-2mm}
\tablecomments{Column (1): source number which is organized by increasing galactic longitude.
         Sources detected by \citet{digel94} and \citet{kerton2003} are marked with $\dagger$ and *. 
	 Columns (2)-(3): Galactic coordinates of the CO emission peak. Columns (4)-(7): results
         of Gaussian fit to the spectra. Column (8): angular area of the complex defined by the
         3$\sigma$ limits. Columns (9)-(10): the heliocentric distance $d$, and the Galactocentric
         distance $R$, respectively; both are derived from the rotation curve of Reid et al. (2014),
         assuming $R_{\sun}$ = 8.34 kpc and $\Theta_{0}$ = 240~\kms.
         Column (11): scale height, $z$, $z=D sin(b)$.
	 Column (12): equivalent radii of the molecular clouds corrected by the beam size of the telescope.
         Column (13): cloud mass calculated using $X$ = 1.8$\times$10$^{20}$~(Dame et al. 2001).}
	 \end{deluxetable}

\begin{figure}
\centering
\includegraphics[angle=0,scale=0.9]{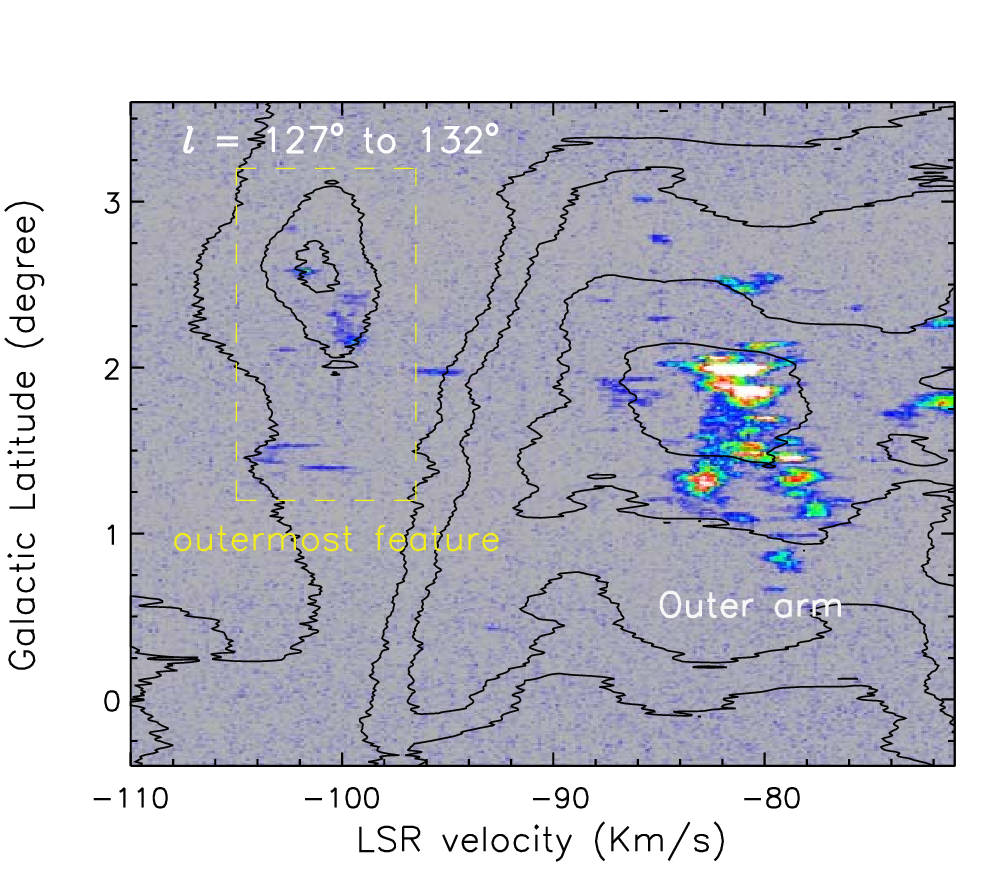}
\caption{Latitude$-$velocity diagrams of CO~(image) and HI~(contours) integrated in
$l$ from 127$^{\circ}$ to 132$^{\circ}$. The dashed box indicates the approximate
boundaries of the outermost feature. The contour levels are 35, 75, 100, 150 and 200~K.
The Outer Arm and the outermost feature are clearly seen as strong, separate features.\label{fig:fig2}}
\end{figure}
\begin{figure}
\centering
\includegraphics[angle=0,scale=0.8]{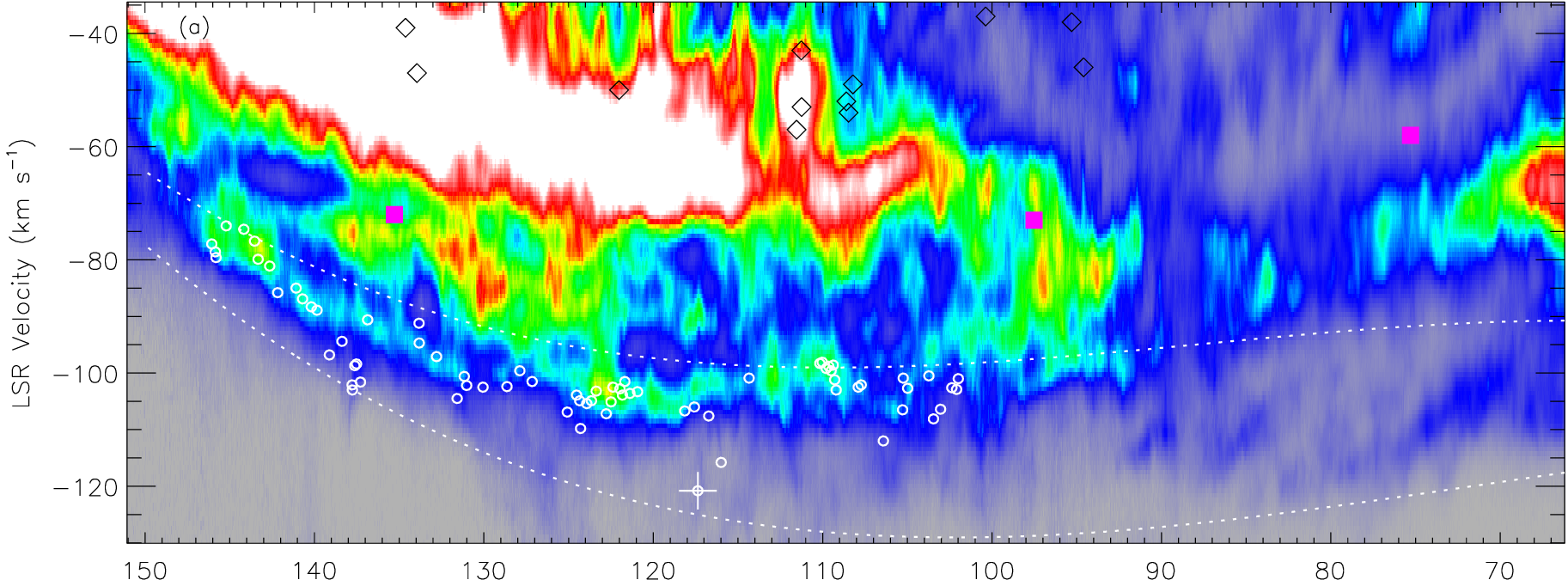}
\includegraphics[angle=0,scale=0.8]{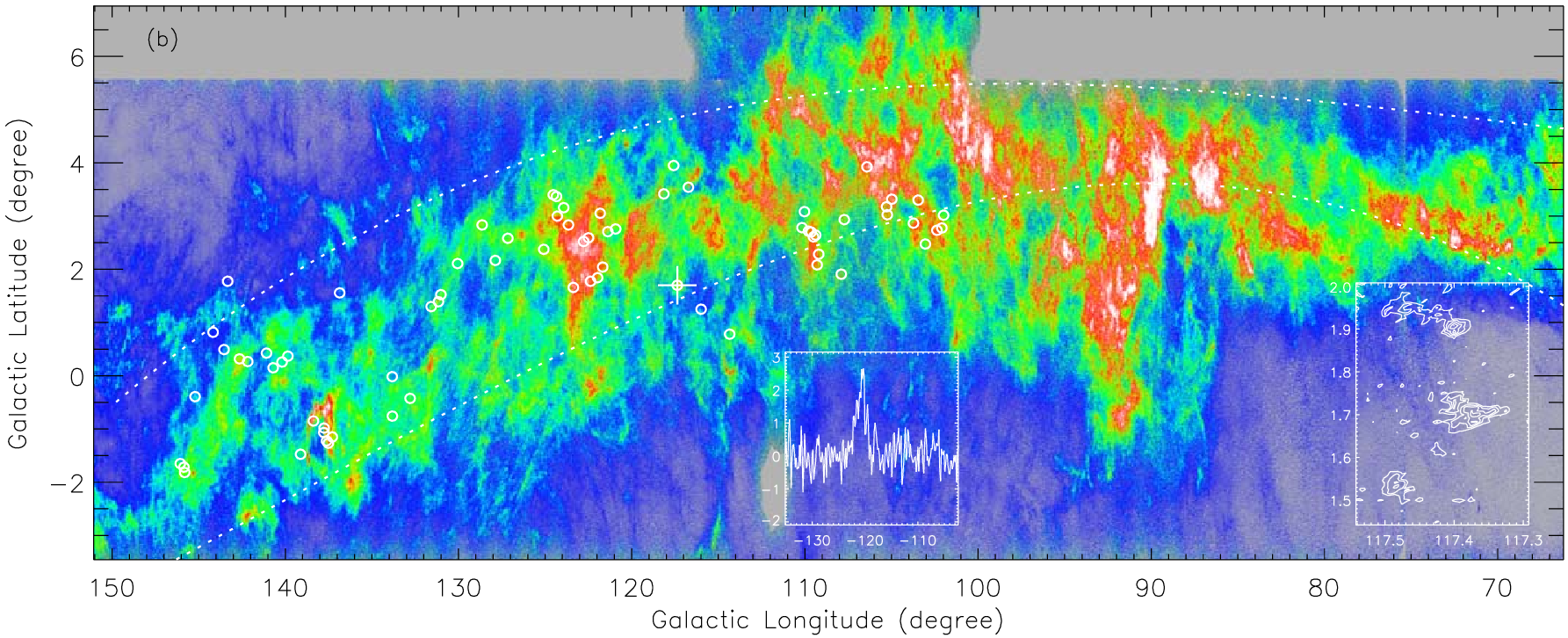}
\caption{(a) Longitude-velocity diagram of HI~(image), integrated
over a window that follows the new arm in latitude. The circles mark the locations of the molecular complexes
summarized in Table~1. The HMSFRs assigned to the Outer and Perseus
arms are marked with squares and diamonds~\citep{reid2014}.
(b) Velocity integrated intensity of HI corresponding to the
new arm. The integrated velocity window is marked by the two dashed lines in
the upper panel. The insets show the velocity$-$integrated CO map and
spectrum of peak emission of molecular clouds 24~(marked with cross).
The lowest contour level is 3$\sigma$.
 \label{fig:fig3}}
\end{figure}
\begin{figure}
\centering
\includegraphics[angle=0,scale=0.73,bb=70 345 479 700]{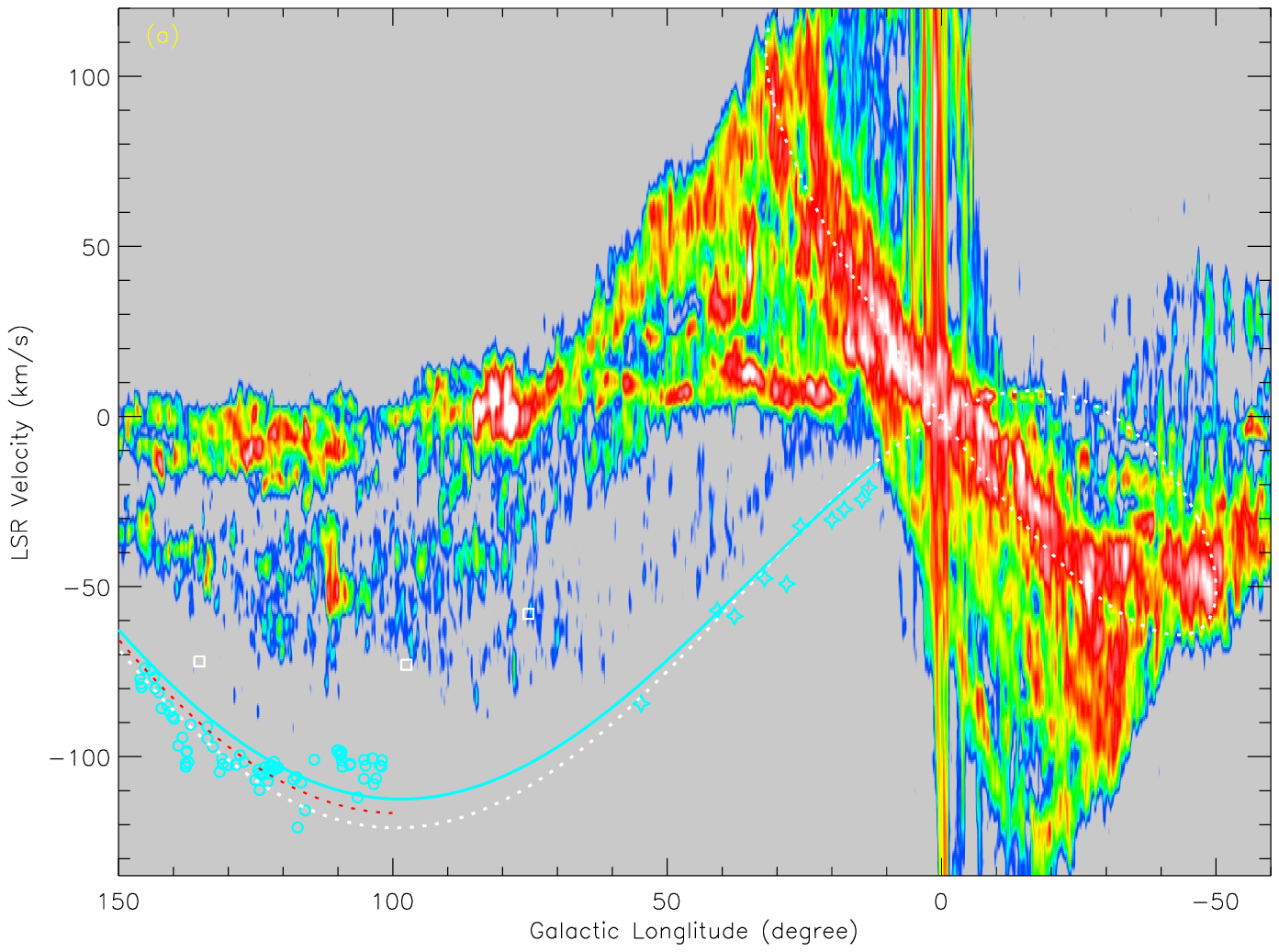}%
\includegraphics[angle=0,scale=0.73,bb=85 345 394 700]{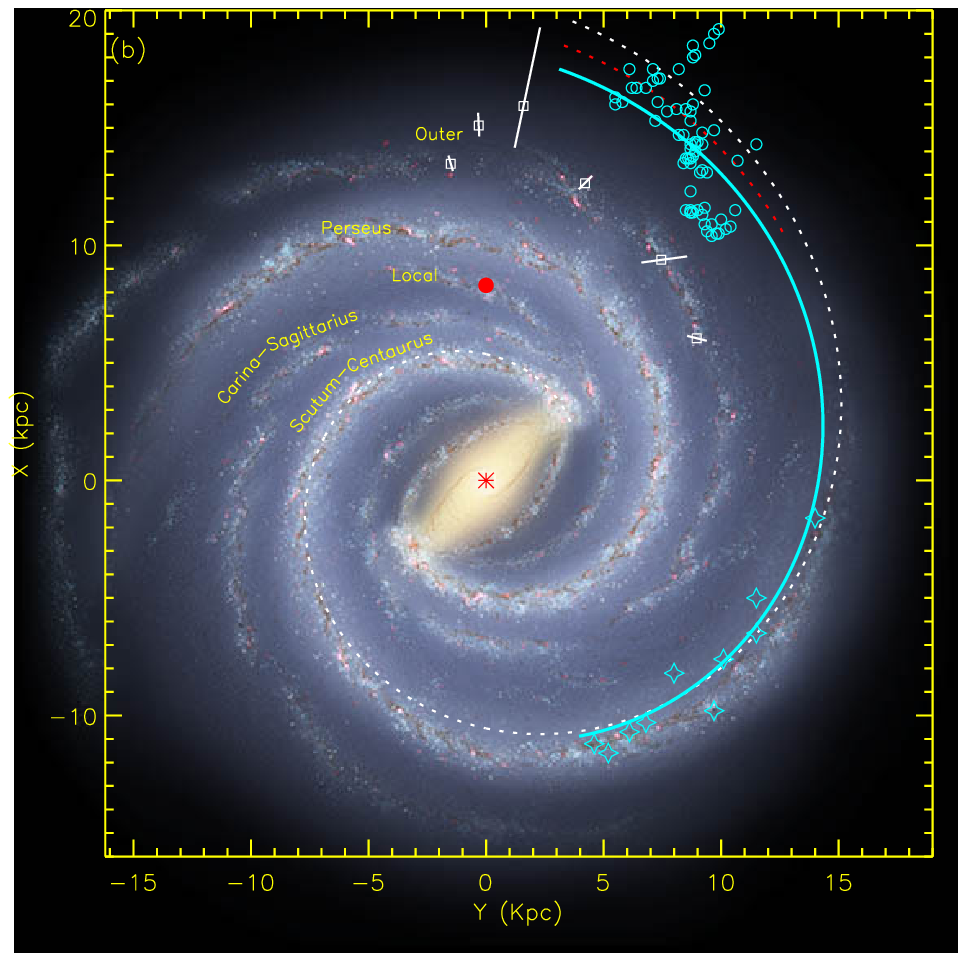}
\caption{Locations of molecular clouds in the far outer Galaxy
superposed on: (a) CO longitude velocity diagram from \citet{dame2001}, and (b) an artist's
conception of the Milky Way~(R. Hurt: NASA/JPL-Caltech/SSC).
The filled stars and circles mark clouds covering Galactic longitudes from
$l$=13$^{\circ}$ to 55$^{\circ}$ detected by \citet{dame2011}, and $l$=100$^{\circ}$ to
150$^{\circ}$ detected by us, respectively.
The squares indicate the locations of HMSFRs associated with the Outer arm.
The white dashed line is a log spiral with a mean pitch angle of 
12$^{\circ}$ that was fit to the Scutum-Centaurus arm in the inner 
Galaxy~\citep{vallee2008,vallee2014}. The red dashed line traces the
log-periodic spiral fitting results to data in 120$^{\circ}$ to 150$^{\circ}$, while the
cyan solid line traces the fitting results to all available data including clouds 
detected by \citet{dame2011}~(see Section 3). \label{fig:fig4}}
\end{figure}
\end{document}